\documentclass[english,twocolumn]{revtex4}
\usepackage[latin1]{inputenc}
\usepackage{graphicx}
\usepackage{amsmath, amssymb,empheq,xcolor} 
\usepackage[small,]{caption}
\usepackage{hyperref}
\makeatletter
\usepackage{babel}
\makeatother

\begin{document}

\title{Photothermal Single Particle Microscopy: Detection of a Nanolens.}
\author{Markus Selmke, Marco Braun, Frank Cichos}
\address{Molecular Nanophotonics Group, Institute of Experimental Physics I, University Leipzig, 04103 Leipzig, Germany}

\date{\today}
\maketitle
Photothermal microscopy \cite{Boyer2002,Lounis2004,Lounis2005,Orrit2010Science} has recently complemented single molecule fluorescence microscopy by the detection of individual nano-objects in absorption. Photothermal techniques gain their superior sensitivity by exploiting a heat induced refractive index change around the absorbing nano-object. Numerous new applications to nanoparticles, nanorods and even single molecules \cite{Hartland2010, Berciaud2008,Rings2010PRL,Waehnert2009,Lounis2009PhoCS,Raduenz2009} have been reported all refering to the fact that photothermal microscopy is an extinction measurement on a heat induced refractive index profile. Here, we show that the actual physical mechanism generating a photothermal signal from a single molecule/particle is fundamentally different from the assumed extinction measurement \cite{Gaiduk2010,Berciaud2006}. Combining photothermal microscopy, light scattering microscopy as well as accurate \textsc{Mie} scattering calculations to single gold nanoparticles, we reveal that the detection mechanism is quantitatively explained by a nanolensing  effect of the long range refractive index profile. Our results lay the foundation for future developments and quantitative applications of single molecule absorption microscopy.

\begin{figure}
	\begin{center}\includegraphics [width=1\columnwidth]{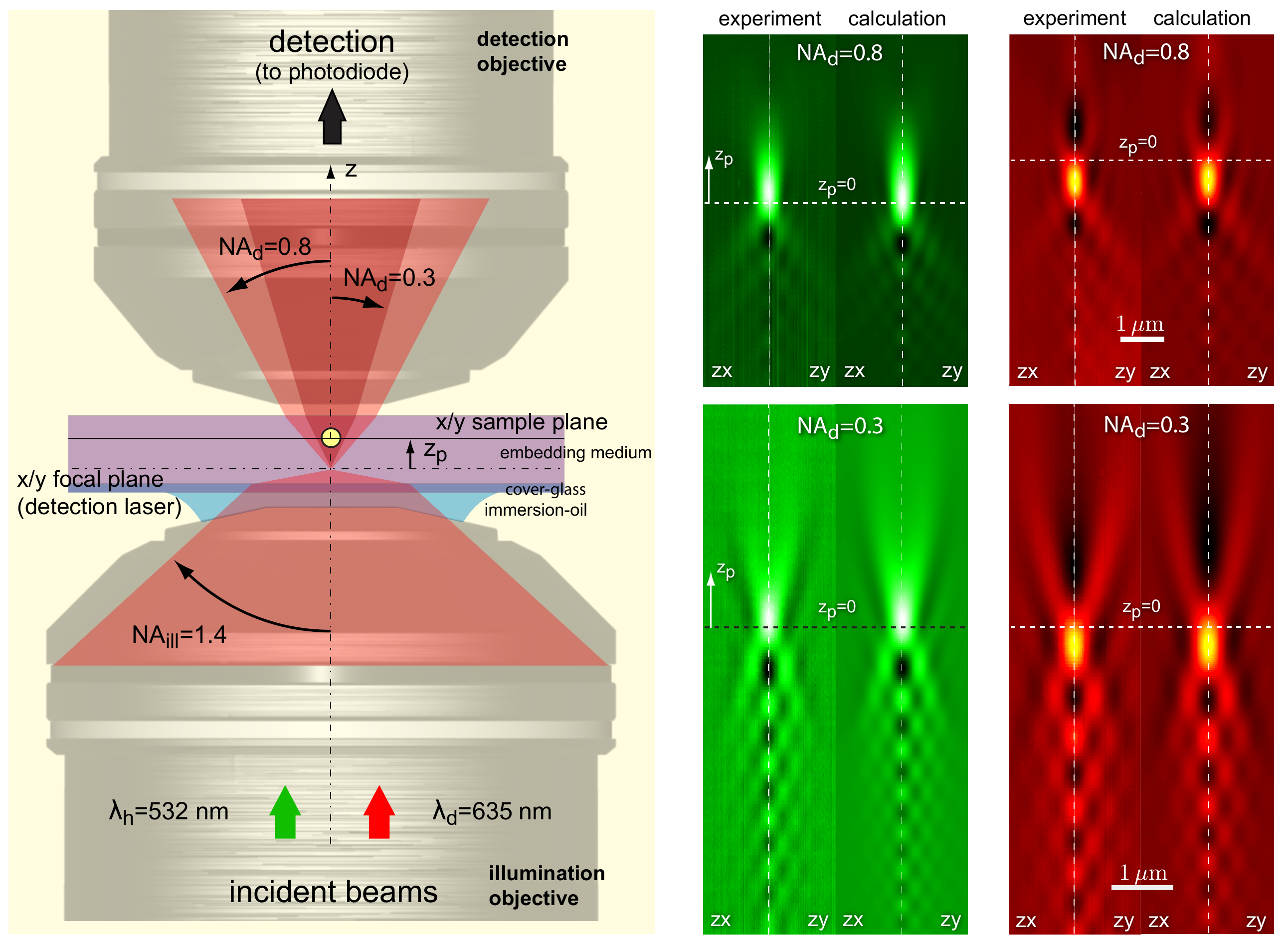}
\end{center}
\caption{{\bf Single gold-nanoparticle light scattering.} (left) Schematic representation of the experimental setup used for the images (sample scanning). Shown is the particle position $z_{p}$, whereas $x_p$ and $y_p$ are the lateral directions (not depicted). (center) Scattering of the heating laser on the gold nanoparticle (AuNP). (right) Scattering of the probe laser on the AuNP ($R=30\,\rm nm$). The top-row images represent the scans for a detection aperture ${\rm NA}_d=0.8$, the bottom row images ${\rm NA}_d=0.3$. The images are grouped experimental/theoretical scans.\label{figure1}}
\end{figure}
Long range interactions play a very special role in physics. Fundamental forces such as the Coulomb force or gravitation are attributed to interaction potentials with inverse distance dependencies manifesting their action even at macroscopic length scales. In this paper we explore the action of another long-ranged field interacting with photons, namely the refractive index profile generated in the solvent surrounding a suspended nanoparticle that releases heat. A typical practical situation is provided by a gold nanoparticle in the focus of a laser beam. Most of the absorbed optical energy is released as heat to the solvent. This creates a temperature field which decays with the inverse distance $r$ from the absorber according to
\begin {equation}
\Delta T\!\left(r\right)=\Delta T_0\frac{R}{r}\,,
\end{equation}
where the temperature rise at the particle surface $\Delta T_0=P_{\rm abs}/4\pi \kappa R$ is resulting from the absorbed optical power $P_{abs}$, the surroundings' heat conductivity $\kappa$ and the particle radius $R$. A corresponding refractive-index profile is thereby established, modifying the unperturbed refractive index $n_m$ of the solvent by (eqn. \ref{eqn:RefracGradient}):
\begin{equation}
n\left(r\right) =n_m + \frac{\mathrm{d}n}{\mathrm{d}T} \,\Delta T\left(r\right) = n_m+\Delta n\frac{R}{r}\,.\label{eqn:RefracGradient}
\end{equation}
This infinite hot lens can be exploited to detect even a minute absorber with a probe laser in an optical microscopy setup with extremely high sensitivity\cite{Orrit2010Science}. However, the infinite size of the lens and the complex spatial structure of the tightly focused, aberrated laser beams cause some conceptual and computational challenges for a quantitative understanding of the signal generation\cite{Berciaud2006}. These are addressed below, where we develop a consistent mathematical formalism to quantitatively explain and analyze such "absorption microscopy" experiments in great detail.

The complexity of the signal generation can already be appreciated when both involved lasers (heating and probe) scatter from a single gold nanoparticle at low incident power, where the heating is still negligibly small (see methods section for experimental details). The transmitted intensity collected by a lens reveals strong interference patterns as it senses the field structure of the aberrated incident laser beams (Fig.\ \ref{figure1}, right). Recording these scattering intensities at different detection numerical apertures emphasizes the importance of the phase relation of scattered and transmitted electric fields \cite{Selmke_elsewhere}.
\begin{figure}
	\begin{center}\includegraphics [width=1.0\columnwidth]{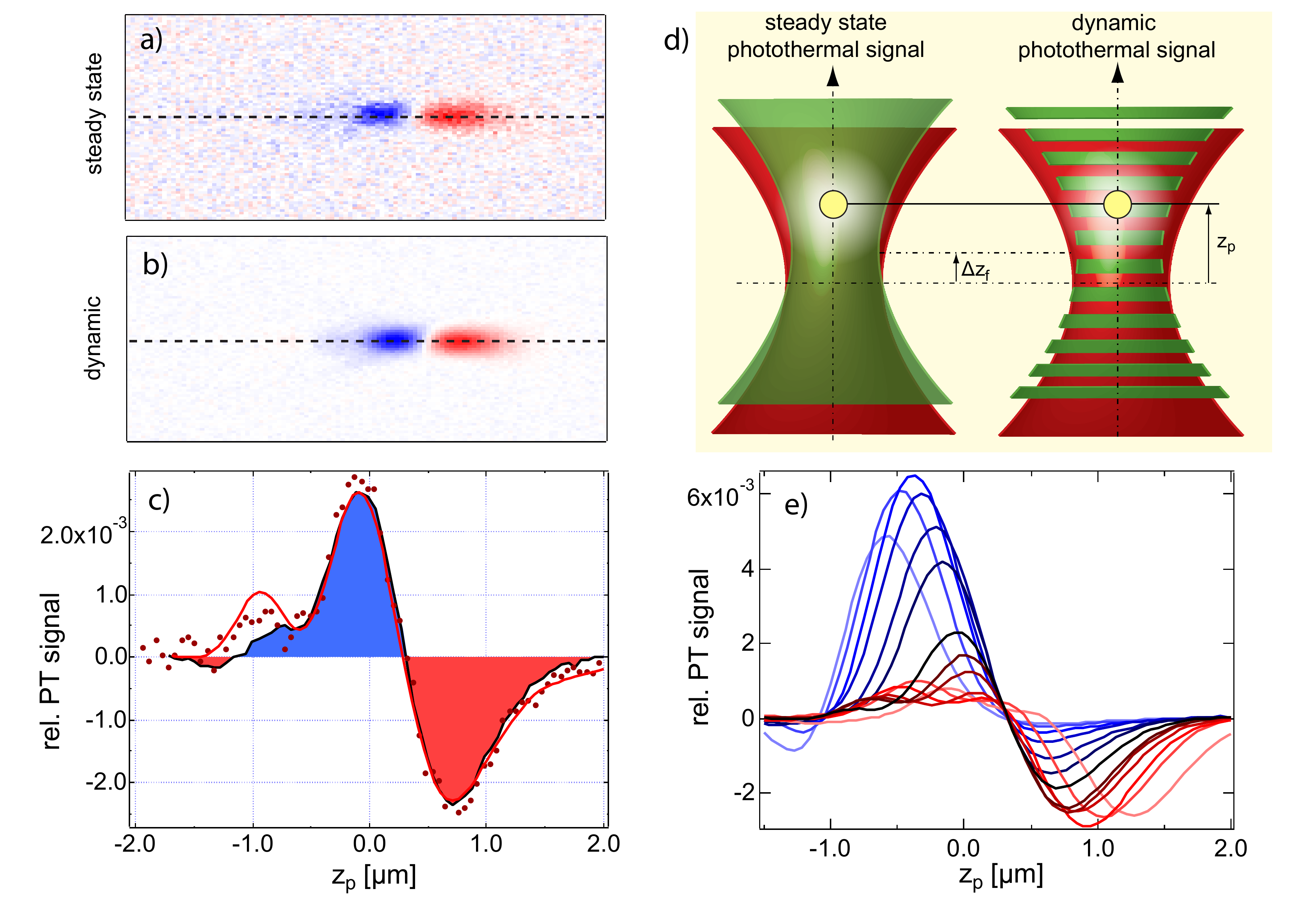}
\end{center}
\caption{{\bf Photothermal signal as a function of heating and probe laser focus displacement.} Difference of hot/cold NP scattering images: a) Static difference image. b) Dynamic PT signal recorded with a lock-in amplifier. c) Comparison of axial PT dynamic (solid black) / static (red dots) / theoretical (solid red) scans. d) Schematic of the defocussing parameter $\Delta z_f$ and the axial particle position $z_p$. e) Experimental PT signal traces for different defocussing parameters $\Delta z_f$. \label{figure2}}
\end{figure}
The interference patterns largely vanish when considering the photothermal signal, which is the difference in the probe laser scattering signal of a heated nanoparticle including the long range refractive index profile and a non-heated particle without the refractive index change. This scattering difference has been evaluated under steady state conditions recording two scattering images as well as dynamically using the photothermal heterodyne technique \cite{Berciaud2006} for an axial focus displacement of both lasers of $350 \, {\rm nm}$ (see Fig.\ \ref{figure2} a and b). Despite the much lower signal to noise ratio of the steady state difference, both signals agree perfectly and reveal a two-lobe structure (Fig.\ \ref{figure2} c). The simplicity of this two-lobe structure suggests a much simpler mechanism than the complex aberrated spatial heating and probe laser scattering intensity distributions put forward. It also unveils that the common assumption of a product point spread function of heating and detection laser is not appropriate. Instead, the two-lobe structure depends sensitively on the displacement of the two laser foci. Whenever the probe laser focus is in front of the refractive index gradient, the detected intensity is decreased, while an increased signal is measured when the probe laser focus is behind the refractive index gradient, as indicated in Figure \ref{figure2} e. This is exactly the action of a diverging lens. While it is in general tempting to consider the refractive index profile as a finite sized scatterer which has to be treated by an appropriate scattering theory, we infer that the missing characteristic length scale of refractive index profile is the source of this intuitive lens-like action. As most materials lower their refractive index with increasing temperature ($\mathrm{d}n/\mathrm{d}T<0$), the temperature field induces a diverging lens with radial symmetry. In fact a more detailed analysis reveals that the problem is equivalent to \textsc{Rutherford} scattering of $\alpha$-particles on a \textsc{Coulomb} potential in which case a classical mechanics treatment of the $\alpha$-particle delivers the same result as a quantum mechanical treatment of the particle \cite{Baryshevskii2004}. While we will detail this fundamental equivalence in a separate paper \cite{PhotonicRutherford2011}, we will stress a complex scattering description to demonstrate that an exact electromagnetic treatment is providing the same picture of a gradient index nanolens. 

\begin{figure}
	\begin{center}\includegraphics [width=1.0\columnwidth]{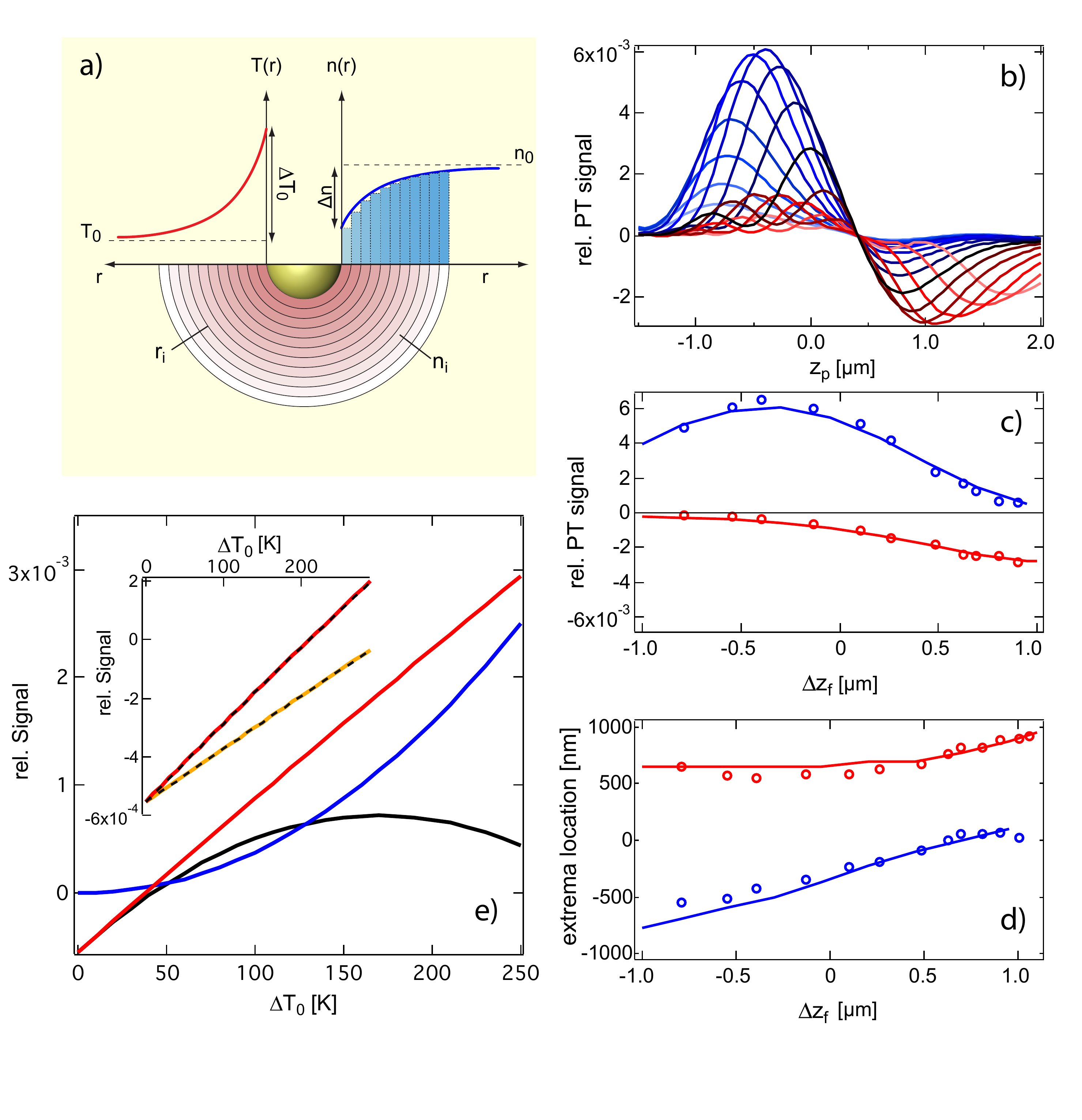}
\end{center}
\caption{{\bf Generalized \textsc{Mie} scattering calculation results of the photothermal signal.} a) Illustration of the temperature profile $T\left(r\right)$, the refractive index profile $n\left(r\right)$ and the discretization of the latter. b) Theoretical photothermal $z_p$ scans for varying foci displacements $\Delta z_f$. c) Peak amplitudes vs. $\Delta z_f$. d) Peak positions vs. $\Delta z_f$. e) Signal decomposition for a $R=10\,\rm nm$ AuNP at the positive lobe $z_p$ position: $P_{\rm ext}$ (black), $P_{\rm sca}$ (blue), $P_{\rm ext}+P_{\rm sca}$ (red). The inset shows $P_{\rm ext}$ and $P_{\rm ext}+P_{\rm sca}$ for $n\left(r\right)=n_m+\Delta n R^2/r^2$ (dashed / orange) and $n\left(r\right)=n_m+\Delta n \exp(-\left(r-R\right)/2R)$ (dashed / red)\label{figure3}}
\end{figure}

This theoretical scattering description has to go beyond common \textsc{Mie} theory as photothermal microscopy employs highly focused laser beams instead of plane waves. Thus a more rigorous \textsc{Mie} description (Generalized \textsc{Lorenz}-\textsc{Mie} Theory, GLMT \cite{Gouesbet1995}) has been extended to account for the axial structure of the signal and to accurately model aberration effects in the focused laser beams \cite{Neves2007,Nasse2010} interacting with the metal particle and the refractive index gradient. The latter one is introduced in a multiple shell approach discretizing the refractive index profile as described by Pe\~na et al. \cite{Pena2009} (see Fig.\ \ref{figure3} a). The detection aperture ${\rm NA}_d$ has been introduced into the GLMT formalism calculating the time-averaged Poynting vector $\langle \mathbf{S}\rangle_t$ integrated over an area $\mathcal{A}_d$ representing an angular detection domain which is determined by the detection aperture (eqn. \ref{eq:poynting}).
\begin{equation}\label{eq:poynting}
P_d = \int_{\mathcal{A}_d} \langle \mathbf{S}\!\left(\mathbf{r}\right) \rangle_t \cdot \,\mathrm{d}\mathbf{A}=\frac{1}{2} \int_{\mathcal{A}_d} \mathcal{R}\left\{\mathbf{E}\!\left(\mathbf{r}\right) \times \mathbf{H}^{*}\!\left(\mathbf{r}\right)\right\} \cdot \,\mathrm{d}\mathbf{A}
\end{equation}
Incorporating all experimental and material parameters (see method and supplement) to evaluate the pure gold nanoparticle scattering, an unprecedented quantitative agreement with the experimental results is unveiled (see Fig. \ref{figure1}). To our knowledge, this is the first quantitative exact calculation of single gold nanoparticle scattering intensity distributions. The calculated scattering images are largely determined by aberrations of the incident laser fields and additional interference structures. The maximum detected scattering intensity and true focus position do not coincide and strongly depend on wavelength. The spatially extended intensity distribution causes the peak intensity to be much lower than for a Gaussian beam having a beam-waist of the corresponding diffraction limit ($\approx 0.61 \lambda / {\rm NA_{\rm ill}}$). Thus, as compared to such a Gaussian beam only one third of the particle temperature is reached.  
The amplitude of the aberration induced interference pattern is determined by the numerical aperture of the detection lens as found in the experiment. The larger this detection aperture, the weaker the interference patterns get as the individual phase differences at different detection angles average out (Fig.\ \ref{figure1} right). This finding confirms that the plane-wave optical theorem is not applicable when collecting signals at finite detection angle and focused illumination \cite{Gouesbet1996opttheorem}. The only way to conclude on the intensity distribution details in the scattering experiment is thus to consider the finite detection angle. While this complex structure determines the shape of the scattering intensity distribution, a contribution to the photothermal signal is also expected \cite{Jurgensen1995,Bialcowski1996} but less obvious. 

We evaluate the photothermal signal as the difference between the probe scattering intensity distribution of particle with the refractive index profile and the distribution without the profile, completely equivalent to the above presented experiments. Accordingly, the theoretical model directly confirms the experimental observations quantitatively. Almost all of the interference structure disappears due to the difference of the two aberrated signals. A two lobe structure is remaining, which exactly corresponds to the experimental observations (Fig.\ \ref{figure2} c). Even the strong dependence of the total signal on the displacement of the two involved laser foci is reproduced validating our theoretical approach (Fig.\ \ref{figure3} b). Thus the exact treatment of the signal by a generalized \textsc{Mie} scattering calculation predicts a lens-like action of the refractive index gradient as well. It demonstrates that a quantitative analysis of photothermal microscopy data in terms of temperatures, absorption cross sections and even in terms of the sign of the thermorefractive coefficient $\mathrm{d}n/\mathrm{d}T$ is possible. This puts photothermal microscopy to a new quantitative level.

The fundamental difference between the commonly assumed extinction process and the mechanism reported here becomes obvious when separating the individual contributions to the signal as done for instance by Gaiduk \cite{Gaiduk2010}. The interaction of a probing field $\mathbf{E}_{\rm pr}$ with the heat induced scatterer is commonly solved through the introduction of an outgoing spherical wave $\mathbf{E}_{\rm sca}$. The detected power $P_d$ then mathematically decomposes, upon inserting $\mathbf{E}=\mathbf{E}_{\rm pr}+\mathbf{E}_{\rm sca}$ (and $\mathbf{H}$) into eqn.\ \ref{eq:poynting}, into three separate integrals \cite{Berg2008} for probe background $P_{\rm pr}$, scattering $P_{\rm sca}$ and extinction $P_{\rm ext}$ powers containing only the probing field $\mathbf{E}_{\rm pr}$, the scattered field $\mathbf{E}_{\rm sca}$ and both in mixed terms, respectively. So far, it has been commonly assumed that solely the latter extinction part contributes to the photothermal signal \cite{Berciaud2006}. This assumption does not hold, independent of the actual gold nanoparticle size thus being valid for single molecule detection as well. A separation of scattering and interference contribution in our theory shows that both parts change in a nonlinear fashion when increasing the temperature of the nanoparticle (see Fig.\ \ref{figure3} e). While up to a temperature rise of about 50 K the interference contribution determines the signal, the importance of the scattering contribution rises quickly. It is the consequence of the inverse distance dependence of the refractive index change, which removes the characteristic length scale from the profile and thus requires a detailed theoretical analysis. This actually becomes obvious when considering a gold particle surrounded by an artificial refractive index profile which decays exponentially or with the inverse distance squared. In both cases, the finite length scale of such a refractive index profile ensures a leading contribution coming from the interference term (Fig.\ \ref{figure3} e inset) increasing linearly with the temperature rise of the particle. 
\begin{figure}
	\begin{center}\includegraphics [width=1\columnwidth]{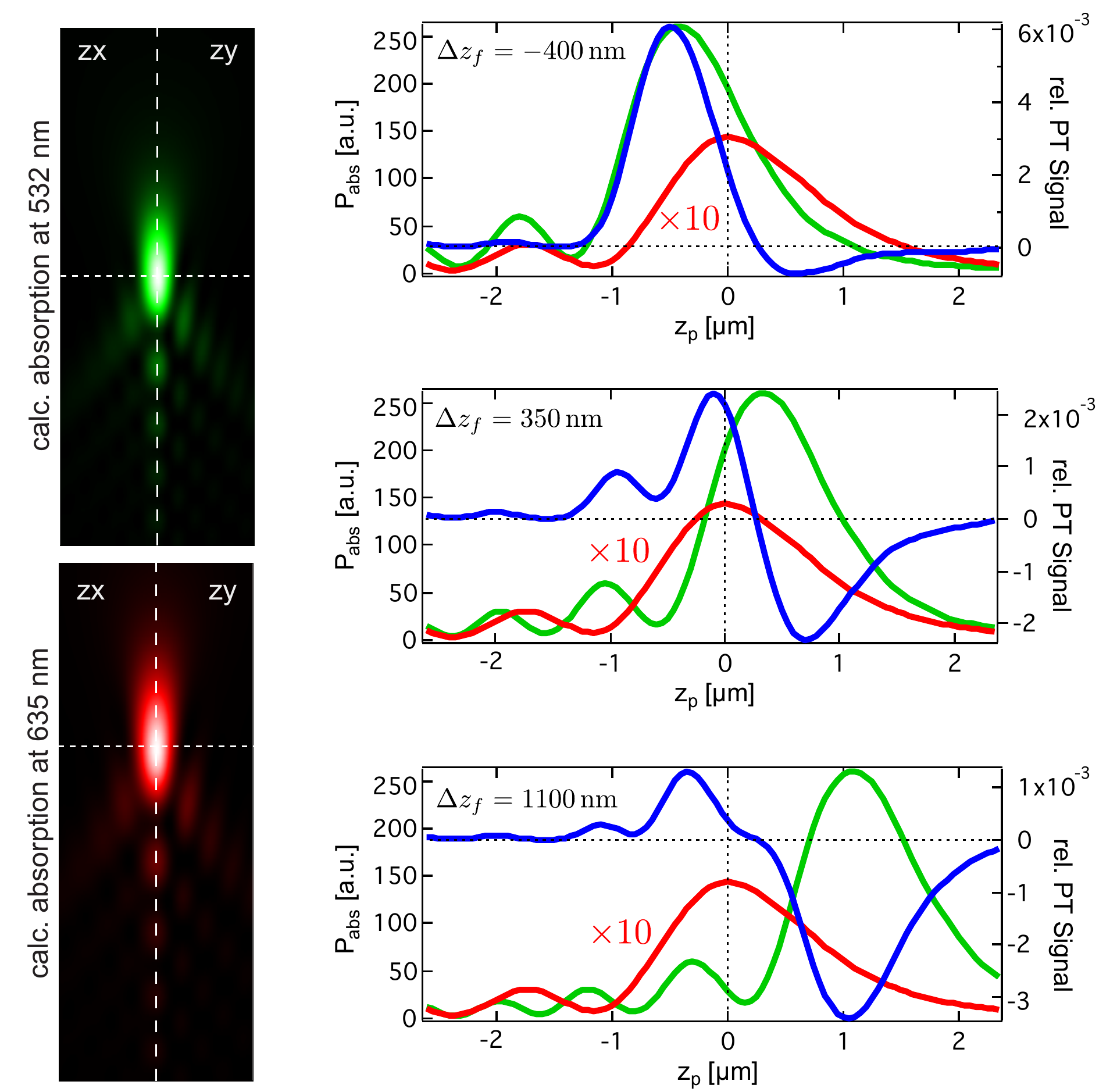}
 	\end{center}
\caption{{\bf Comparison photothermal and fluorescence signal.} Left column: Calculated scans of the absorbed power $P_{\rm abs}$ ($\propto$ point spread function, $\left|\mathbf{E}\right|^2$) for the heating laser (top) and the detection laser (bottom). Right column: Representative scenarios for the relative alignments of the two aberrated lasers: (top) maximal positive photothermal signal, (center) symmetric signal and (bottom) maximal negative signal. Plotted are the heating beam (green, fluorescence) and probe beam (red) intensities as well as the resulting photothermal signals (blue) vs. the axial particle coordinate $z_p$.\label{figure4}}
\end{figure}
Therefore, the missing length scale of the refractive index change is responsible for both scattering and interference contributions to the signal, which have to be treated explicitly to understand the signal generation mechanism. The overall quality of our theoretical and experimental results is best seen in Figure \ref{figure3} c and d, plotting the maximum and minimum photothermal signal values as well as their positions as a function of the axial focus displacement $\Delta z_{f}$. They show an unprecedented agreement of experiment (markers) and theory (lines) without the inclusion of fitting parameters. 
From the above consideration of different refractive index profiles it is evident, that a calibration of the photothermal signal for the measurement of absorption cross-sections on arbitrarily shaped objects of sizes comparable to the lateral focus-extent (e.g. single-walled carbon-nanotubes \cite{Berciaud2008,Hartland2010}) will fail as the general contribution of scattering and interference will change. In the case of point-like absorbers, however, there is a quantitative simplification of the generalized \textsc{Mie} theory applied here. In fact the lens-like action can be treated in a geometrical optics framework due to its fundamental mathematical equivalence to \textsc{Rutherford} scattering as we will show elsewhere\cite{PhotonicRutherford2011}.

Finally, the contribution of aberrations to the photothermal signal can be rationalized. The photothermal signal generated by a non-aberrated probe beam is vanishing if the particle is in the focus of the probe laser. This behavior is expected for a simple lens. The aberrated probe beam, however, leads to a finite signal at the particle position. This can be understood in terms of a lens-like action as well. While a non-aberrated intensity distribution shows a zero crossing of the signal at the particle position (not shown), the aberrated probe beam leads to a finite signal at the particle position. While a non-aberrated beam is symmetric to the lens if directly focused to the center of the lens, the aberrated beam is not (Fig.\ \ref{figure4}, left). The additional interference maxima act like additional foci displaced with respect to the photothermal lens position and thus cause a photothermal signal. As a consequence, the displacement of particle fluorescence excited by the heating laser and photothermal signal will depend on the aberration of the probe beam and not simply on the Gouy phase \cite{Hwang2007}. We have calculated three limiting cases, where either the positive lobe or the negative lobe is maximum or both lobes show the same photothermal signal magnitude (Fig.\ \ref{figure4}, right). In the first case, photothermal signal maximum and fluorescence signal maximum would be displaced by about 100 nm (top). In the second case, the fluorescence signal is almost exactly at the position of the photothermal signal minimum (center) while in the last case the zero-crossing of the photothermal signal coincides with the fluorescence maximum (bottom). Therefore the commonly found displacements of fluorescence and photothermal signal \cite{Hartland2010,Orrit2010PCCP} in axial direction can be well explained by the presented theoretical approach. 

In summary, our results demonstrate that the long range refractive index change generated by a single heated nano-object in photothermal microscopy act as a nano lens. The understanding of this signal generation mechanism establishes photothermal microscopy as a quantitative technique to determine absolute absorption cross sections and delivers a framework for new applications of this technique. As a direct consequence, new experimental techniques such as twin-focus photothermal correlation spectroscopy similar to the well established dual focus fluorescence correlation techniques \cite{Dertinger2007} or super-resolution absorption microscopy methods can be developed. We expect that the understanding of the photothermal signal generation will pave the way for further improvements of this technique beyond the current level of sensitivity.

\subsection{Sample preparation.}
Samples were prepared by spin-coating a polymer layer (Sylgard 184, about $15\,\mathrm{\mu m}$ thickness) on top of a glass-cover slide. AuNPs (BBI International) with a diameter of $60\,\mathrm{nm}$ were deposited on the polymer film. The particles were covered with a second Sylgard layer (also about $15\,\mathrm{\mu m}$ thick) to embed the particles in a homogeneous matrix. The absence of a close-by interface ensures a radially symmetric temperature field around the particle.

\subsection{Photothermal microscopy measurements.}

\begin{figure}
	\begin{center}\includegraphics [height=0.6\columnwidth]{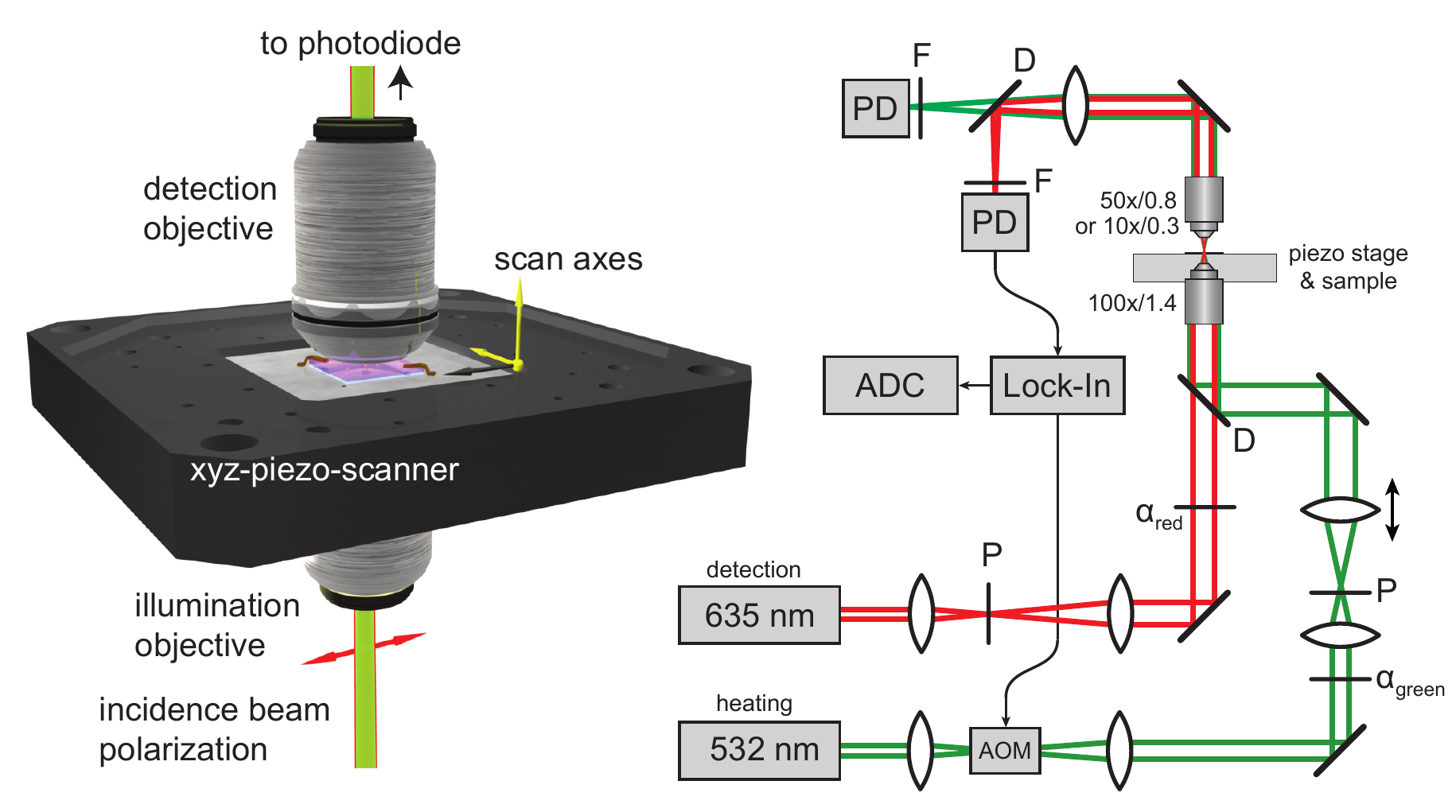}
\end{center}
\caption{\label{cap:dist}Principle scheme of the experimental setup. PD: photo-diode, P: pinhole, D: dichroic mirror, F: filter, AOM: accusto-optic modulator, $\alpha$: variable ND filter, ADC: Adwin analog digital converter.}
\end{figure}

The experimental setup for single particle light scattering and photothermal measurements is based on a home-built confocal sample-scanning microscopy setup using two laser sources. A DPSS laser (Coherent, Verdi) with $\lambda_{h}=532\,\rm nm$ is used to heat the gold particles and a second laser source at $\lambda_{d}=635\,\rm nm$ (Coherent ULN laser diode) probes the local refractive index changes. Both beams are focused into the sample by the same objective lens (Olympus 100x/1.4NA) and are collected above the sample by a second objective (Zeiss 10x/0.3NA or Olympus 50x/0.8NA), which is adjusted to image the probe focus to infinity. The sample is moved by a piezo-scanner (PI). The resulting parallel beam is focused onto two photodiodes (Thorlabs, PDA36A-EC) after passing appropriate filters (no pinhole). To allow for a low noise detection of the photothermal signal, the heating beam is modulated with a frequency $\Omega= 300\,\mathrm{kHz}$ and the probe signal change is detected at this reference frequency with a lock-in amplifier (Signal Recovery 7280 DSP) with a time constant of $\tau_{\rm li}=1\,$ms. The resulting lock-in signal is recorded by a A/D converter (Adwin-Gold, J\"ager Messtechnik) 300 times for each recorded pixel (1ms/pixel).
The relative photothermal signal $\Phi$ corresponds to the relative modulation amplitude at the photodiode
\begin{equation}
\Phi=\frac{\Delta V}{\langle V\rangle}
\end{equation}

\subsection{Single particle light scattering measurements.}
Single particle light scattering has been carried out in the same setup as the photothermal measurements. To do so, the intensity of heating and probe laser have been diminished that no notable effects of varying incident laser power can be found in the experiments. The laser intensities of both wavelength were recorded independently with two photodiodes (Thorlabs, PDA36A-EC) without intensity modulation and lock-in detection. All scattering signals have been normalized to the background intensity. 

\subsection{\textsc{Mie} scattering calculations.}
Mie scattering calculations were carried out with a modified C-code. The multishell scatter coefficients $a_n^{L+1}$ and $b_n^{L+1}$ were provided by ref. \cite{Pena2009}, and are referred to in the supplemented material as $a_n$ and $b_n$. The modifications include the numerical evaluation of beam shape coefficients (supplement) as well as the integration of the \textsc{Poynting}-vector across a finite azimuthal collection-angle range (supplement eqn.\ 5-8). The C-code was then interfaced to WaveMetrics Igor Pro 6.06 to create the scans and images.

\subsection{Parameters used for the calculations in Figure \ref{figure1}}
The following parameters were used for the calculation of the imaging (subscript d and h denote detection and heating parameters, see supplement for notation): Particle parameters: $R=30\,{\rm nm}$, $n_{\rm Au}({\lambda_h})=0.516+2.23 i$ and $n_{\rm Au}({\lambda_d})=0.175 + 3.46 i$ \cite{Johnson1972}. Beam and objective parameters: Detection laser: $P_{{\rm PM}, d}=250\mu$W, $\lambda_d=635\,\rm nm$, heating laser: $P_{{\rm PM}, h}=25\mu$W, $\lambda_h=532\,\rm nm$, effective illumination objective focal length $f=1.8$mm, overfilling factors $f/\omega_{d}=\gamma_d=1.125$ and $f/\omega_{h}=\gamma_h=1$. Sample parameters: refractive indices: $n_0=1$, $n_{1,h}=1.525$, $n_{1,d}=1.505$, $n_1^{*}=1.514$, $n_2=n_m=1.46$, $\mathrm{d}n/\mathrm{d}T=-3.6\times 10^{-4}{\rm K}^{-1}$, $\kappa=0.15\,{\rm W m}^{-1}{\rm K}^{-1}$, $k_0 d^{*}=720$, $d=15\,\mu\rm m$, ${\rm NA}_{\rm ill}=1.4$, ${\rm NA}_d=0.3$ or ${\rm NA}_d=0.75$. Discretization parameters: number of angles for numerical angular signal integration: $N_{\theta}=400$, number of layers for multi-shell scatter-coefficients $a_n$ and $b_n$: $N_{L}=370$, layer spacing: $r_j=R+\Delta r j^{1.4}$ with $j=1,\dots,N_L$ and $\Delta r=R/50$. Number of angles for the beam shape coefficients: $N_\alpha=6000$, $m_{\rm max}=30$, $n_{\rm max}$ given by scattnlay. \newline
The bottom row of the Fig.\ \ref{figure4} shows maps of the absorbed power $P_{\rm abs}$. In case of a point-absorber these maps corresponds to the point-spread-functions (PSFs). These maps yield the following effective gaussian parameters: $\omega_{0,d}=281$ nm, $z_{R,d}=570$ nm, $\omega_{0,h}=233$ nm, $z_{R,h}=468$ nm.

\subsection{Temperatures and Cross-Sections from the calculations for Figure \ref{figure2}}
The power measured with the power-meter and converted to peak-to-peak power was $P_{\rm PM}=100\,\mu{\rm W}$ (see supplement). The power-meter aperture was the same as the microscope objective back-aperture, i.e. $c_{\rm PM}=1$. The Gaussian GLMT calculation gives: $\Delta T_0^G=96\,{\rm K}$.
With the expressions for $c_{\rm aberr}$ and $c_G$ (see supplement) we find:
$c_{\rm aberr}=0.384$, $c_{T,h}=0.86$, $c_G=1.225$. The absorption-cross-section was found to be $\sigma_{\rm abs}^G\left(z_p=0\right)= 1.15\times 10^{4}\,{\rm nm}^2$ which is close to the \textsc{Mie}-cross section ($g_n=1$): $\sigma_{\rm abs}^{\rm Mie}= 1.16\times 10^{4}\,{\rm nm}^2$.
The exact calculation yields $\Delta T_0^{E}=95\,{\rm}K$ which is obtained with $\sigma_{\rm inc}^{\rm ill}$ and $\sigma_{\rm abs}^{E}$ (see supplement).
$\sigma_{\rm abs}^E=2.56\times 10^{-7}\,{\rm m}^2$, $\sigma_{\rm inc}^{\rm ill}=4.16\times 10^{-6}\,{\rm m}^2$, $\sigma_{\rm inc}^{d}=1.65\times 10^{-6}\,{\rm m}^2$.

%\bibliographystyle{apsrev}
%\bibliography{MyLibrary10}

\end{document}

\pagestyle{headings}
\maketitle
\vspace{1cm}

\small
\section{Generalized Lorenz Mie Theory}
The Generalized \textsc{Lorenz}-\textsc{Mie} Theory (GLMT) \cite{Gouesbet1995,Gouesbet1988} provides the total Electromagnetic fields $E^t$ and $H^t$ which solve the situation of the incident electric field (denoted by $E^i$ and $H^i$ in the absence of any scatterer) upon a spherically symmetric (possibly multilayered) scattering center. The application of the \textsc{Bromwhich}-formalism to this situation was introduced by G. Gouesbet et al.\ and is here generalized for the purpose of finite collection-angle flux-integration. The field expressions, \textsc{Poynting}-vector expressions and far-field limits of the appropriate Bessel-functions involved were checked and adopted from \cite{Gouesbet1988}. For the calculation of the hot-particle scattering including the refractive index profile, the scattering coefficents $a_n$ and $b_n$ were exchanged for the multilayer coefficients $a_n^{L+1}$ and $b_n^{L+1}$ \cite{Pena2009}.

\section{Integrated Fluxes}
\begin{figure}[h]
  \centering
 	\includegraphics[scale=0.7]{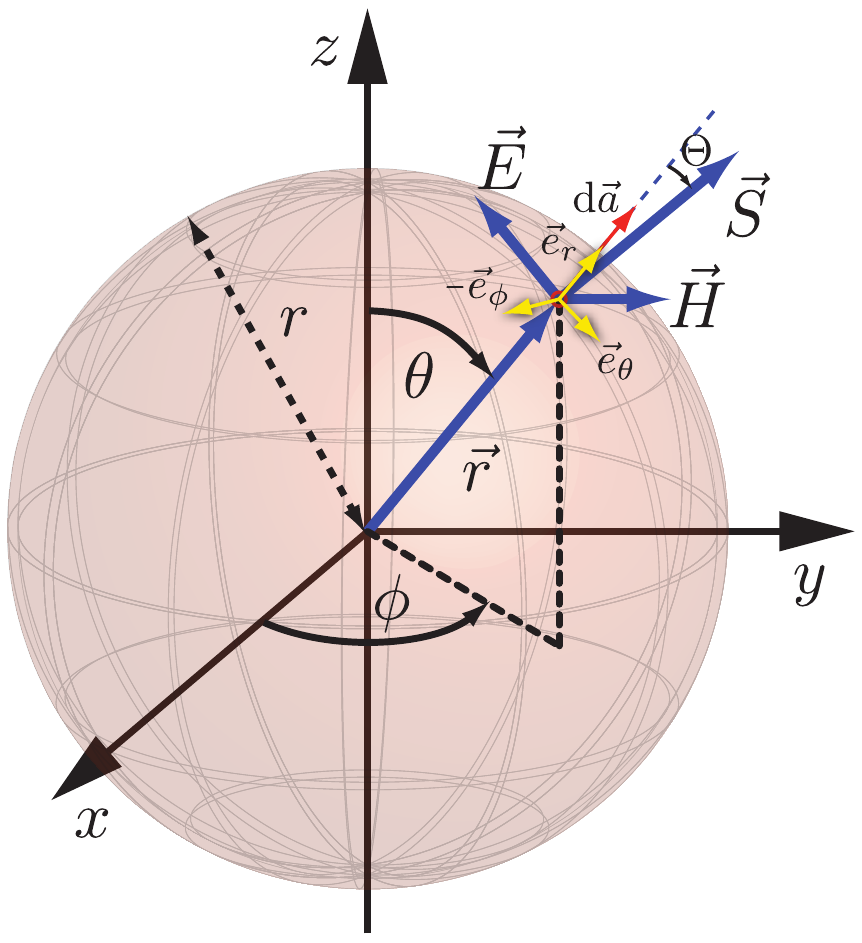}
  \caption{Fields and coordinates}
  \label{fig:Geom}
\end{figure}

To obtain the detected power one may calculate the integrated \textsc{Poynting}-vector of the total EM-field through a spherical surface (see Fig.\ \ref{fig:Geom}) at infinite distance ($r\rightarrow \infty$), i.e. $P_d=\int\mathbf{S}^{t}\cdot \mathrm{d}\mathbf{A}=\int S^{t}_{\perp}\mathrm{d}A$ with $2\langle S^{t}_{\perp}\rangle=\mathfrak{R}\left(E_\theta^t H_\phi^{t*}-E_\phi^t H_\theta^{t*}\right)$, where $\mathfrak{R}\left(z\right)$ denotes the real part of the complex number $z$. Also, $\mathrm{d}\mathbf{A}=\mathbf{n}\,\mathrm{d}A$ with $\mathrm{d}A=r^2 \sin\theta \mathrm{d}\theta \mathrm{d}\phi$ being the surface element in spherical coordinates and $\mathbf{n}$ being the surface normal vector. The integrand may be mathematically decomposed into three parts analogously to the \textsc{Mie}-scattering treatment:
\begin{align}
2\,\langle S^t_{\perp}\rangle =&\mathfrak{R}\left(E_\theta^i H_\phi^{i*}-E_\phi^i H_\theta^{i*}\right)+\mathfrak{R}\left(E_\theta^s H_\phi^{s*}-E_\phi^s H_\theta^{s*}\right)+\nonumber\\
&\mathfrak{R}\left(E_\theta^i H_\phi^{s*}+E_\theta^s H_\phi^{i*}-E_\phi^i H_\theta^{s*}-E_\phi^s H_\theta^{i*} \right)
\end{align}
Energy conservation dictates that $-P_{\rm abs}=\int_{4\pi}\mathbf{S}^{t}_{\perp}\cdot \mathrm{d}\mathbf{A}=P_{\rm inc}^{4\pi}+P_{\rm sca}^{4\pi}+P_{\rm ext}^{4\pi}$, where on the right-hand side the negative sign is introduced to the notation to give the usual physical interpretation of the integral, i.e. $P_{\rm abs}$ being the absorbed power by the sphere. For this expression of energy conservation to hold, one always has $P_{\rm ext}^{4\pi}<0$, while $P_{\rm abs}$, $P_{\rm sca}^{4\pi}>0$. On the other hand, $P_{\rm inc}^{4\pi}=0$, which means that the incident field integral describes a flux into and out of the integration-sphere and a non-absorbing medium is assumed, i.e. $\mathfrak{I}\left(n_m\right)=0$ where $\mathfrak{I}\left(\right)$ denotes taking the imaginary part. Now, let $P$'s denote the integrals over the collected finite angular domain, then $P_d=P_{\rm inc}+P_{\rm sca}+P_{\rm ext}$ with the physically non-individually detectable components given by

\begin{align}
P_{\rm inc}&=\int_{0}^{2\pi}\!\!\!\int_{0}^{\theta_m}\!\frac{1}{2}\mathfrak{R}\left[E_\theta^i H_\phi^{i*}-E_\phi^i H_\theta^{i*}\right]\mathrm{d}A\\
P_{\rm sca}&=\int_{0}^{2\pi}\!\!\!\int_{0}^{\theta_m}\!\frac{1}{2}\mathfrak{R}\left[E_\theta^s H_\phi^{s*}-E_\phi^s H_\theta^{s*}\right]\mathrm{d}A\\
P_{\rm ext}&=\int_{0}^{2\pi}\!\!\!\!\int_{0}^{\theta_m}\!\frac{1}{2}\mathfrak{R}\left[E_\theta^i H_\phi^{s*}+E_\theta^s H_\phi^{i*}-E_\phi^i H_\theta^{s*}-E_\phi^s H_\theta^{i*} \right]\!\mathrm{d}A
\end{align}

\begin{figure}[h]
  \centering
 	\includegraphics[scale=0.48]{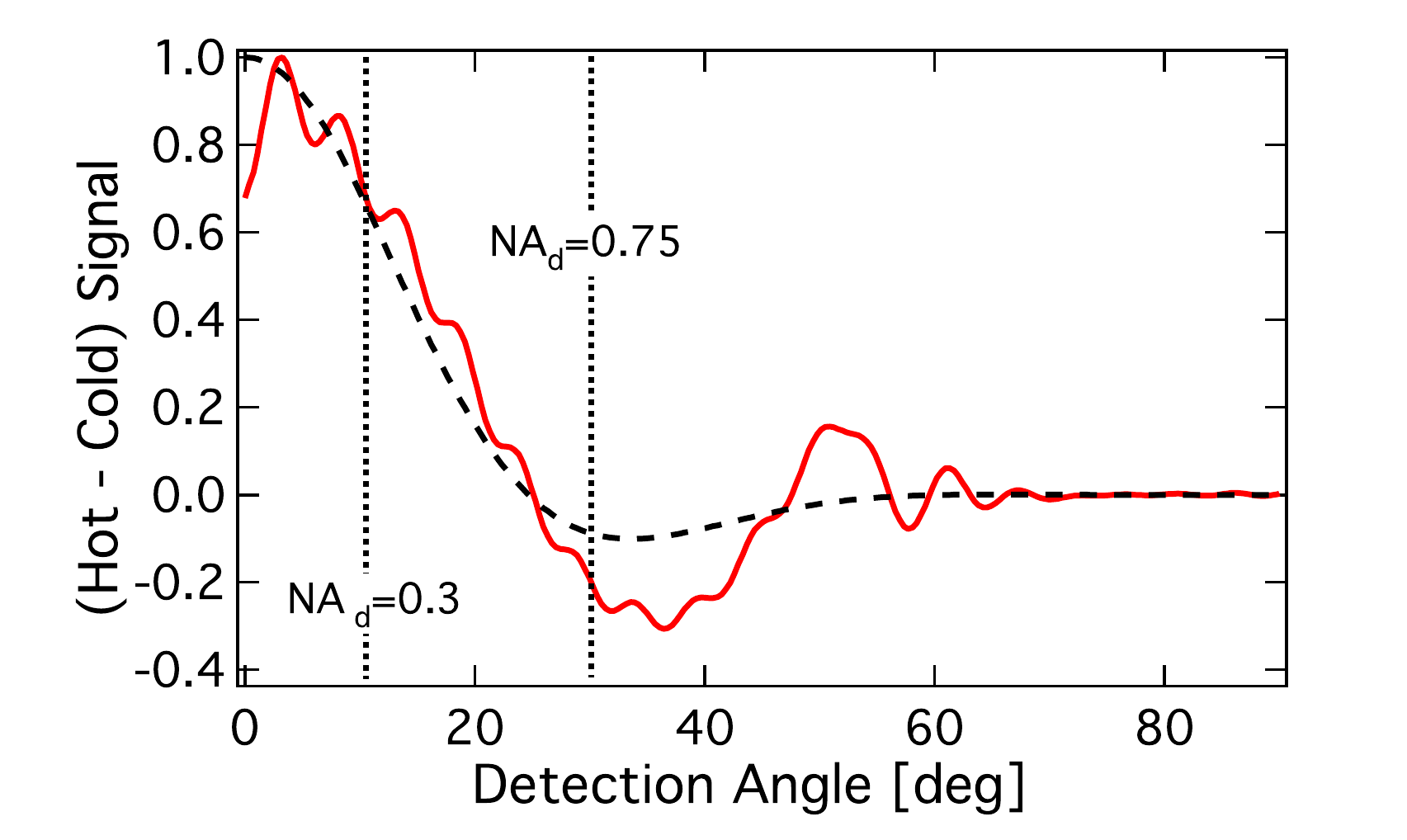}
  \caption{Normalized angular signal spectrum $P_d\left(\theta\right)/\sin\left(\theta\right)$ for $z_p=-z_R$. The vertical lines correspond to the angles $\arcsin\left({\rm NA}_d/n_m\right)$ for ${\rm NA}_d=0.3$ and $0.75$. The solid red line is the signal spectrum for the exact description, while the dashed line corresponds to an on-axis Gaussian beam calculation.}
  \label{fig:Angle}
\end{figure}

The interference and scatter integrals ($P_{\rm ext}$ and $P_{\rm sca}$) were computed with \texttt{Mathematica} leading to the expressions (\ref{eqnOffIS}) - (\ref{eqnM}). Integration over the entire polar angle of $P_{\rm ext, sca}$ with the help of the orthogonality relations for Legendre-Polynomials recovers the expression for the extinction and scatter cross-sections (\cite{Gouesbet1988}, eqn. (\ref{sigmaAbs})), i.e.\ $P_{\rm ext}|_{\theta_m=\pi}=P_{\rm ext}^{4\pi}=-\sigma_{\rm ext} I_0$ and $P_{\rm sca}|_{\theta_m=\pi}=P_{\rm sca}^{4\pi}=\sigma_{\rm sca} I_0$. The intensity $I_0$ (irradiance) to use in these definitions depends on the normalization used in the description of the incident field, i.e. the beam shape coefficients (BSCs) used. In case of the Gaussian GLMT (section \ref{GaussGLMT}), $I_0=2P_0/\pi \omega_0^2$ refers to the focus-peak intensity (and the corresponding cross-sections will have a superscript $^{G}$, i.e.\ $\sigma^G_{\rm abs}$ etc.), while for the exact GLMT $I_0$ refers to the on-axis intensity of the collimated gaussian beam illuminating the microscope objective back aperture (the corresponding cross-sections will have a superscript $^{E}$, i.e. $\sigma^E_{\rm abs}$ etc.). The differential cross-sections are:

\begin{align}
\sigma_{\rm sca}\left(\theta_m\right)&=\sigma_{{\rm sca},1}\left(\theta_m\right)+\sigma_{{\rm sca},2}\left(\theta_m\right)\label{eqnOffIS}\\
\sigma_{{\rm sca},{1 \atop 2}}\left(\theta_m\right)=&\, \frac{2\pi}{k^2}\int_{0}^{\theta_m} \sum_{n=1}^{\infty}\sum_{m=-n}^n \left(X_{1 \atop 2}\right)_{n}^m \sum_{n'=n'_{|m|}}^{\infty} \left(X_{1 \atop 2}^{*}\right)_{n'}^m \sin\theta \mathrm{d}\theta\nonumber\\
\sigma_{\rm ext}\left(\theta_m\right)&=-\mathfrak{R}\left(\sigma_{{\rm ext},1}\left(\theta_m\right)+\sigma_{{\rm ext},2}\left(\theta_m\right)\right)\label{eqnOffIS2}\\
\sigma_{\rm {ext},{1 \atop 2}}\left(\theta_m\right)=&\, \frac{2\pi}{k^2}\int_{0}^{\theta_m} \sum_{n=1}^{\infty}\sum_{m=-n}^n \left(Y_{1 \atop 2}\right)_{n}^m \sum_{n'=n'_{|m|}}^{\infty} \left(X_{1 \atop 2}^{*}\right)_{n'}^m \sin\theta \mathrm{d}\theta\nonumber
\end{align}

with $2\pi/k^2=\lambda_d^2/\left(2 \pi n_{m}^2\right)$ and $n'_{|m|}=\textnormal{max}(1,|m|)$. These expressions have been evaluated numerically. The matrix elements $\left(M_{i,j}\right)_n^m$ will be introduced for convenience and calculated for each angle $\theta$ within the sums (and prior to their evaluation) which helped to speed up the numerical calculations:
\begin{align}
\left(Y_{1 \atop 2}\left(\theta\right)\right)_n^m &=\,\, N_n \!\left[\,{i g_{n,TE}^m \atop g_{n,TM}^m} \tau_n^{|m|}\,+\,{m g_{n,TM}^m \atop i m g_{n,TE}^m} \Pi_n^{|m|}\right]\\
&=\left(M_{{1,1 \atop 2,1}}\left(\theta\right)\right)_n^m+\left(M_{{1,2 \atop 2,2}}\left(\theta\right)\right)_n^m\nonumber\\
\left(X_{1 \atop 2}\left(\theta\right)\right)_n^m &=\,\, N_n \!\left[a_n{m g_{n,TM}^m\Pi_n^{|m|} \atop g_{n,TM}^m\tau_n^{|m|}}\!+b_n {i g_{n,TE}^m \tau_n^{|m|} \atop  i m g_{n,TE}^m \Pi_n^{|m|}}\right]\label{eqnM}\\
&= a_n  \left(M_{1,2 \atop 2,1}\left(\theta\right)\right)_n^m + b_n  \left(M_{1,1 \atop 2,2}\left(\theta\right)\right)_n^m\nonumber
\end{align}

where $i=\sqrt{-1}$ and $N_n=\left[2n+1\right]/\left[n\left(n+1\right)\right]$. The angular functions $\Pi_n^{|m|}$ and $\tau_n^m$ are defined in section \ref{secFunc}. The physically detectable angular power distribution (see Fig.\ \ref{fig:Angle}) is obtained during the numerical evaluation of the integral in the above expressions of eqns.\ (\ref{eqnOffIS}) and (\ref{eqnOffIS2}), i.e.\ $P_d\left(\left[\theta,\theta+\mathrm{d}\theta\right]\right)/I_0=\sigma_{\rm sca}\left(\theta+\mathrm{d}\theta\right)+\sigma_{\rm ext}\left(\theta+\mathrm{d}\theta\right)-\sigma_{\rm sca}\left(\theta\right)-\sigma_{\rm ext}\left(\theta\right)$. A constant incident beam offset is neglected here.

The relation between the powers and the cross-sections are: $P_{\rm sca}^{4\pi}=\sigma_{\rm sca}\left(\pi\right) I_{0}$ and $P_{\rm ext}^{4\pi}=-\sigma_{\rm ext}\left(\pi\right)I_{0}$ with $\sigma_{\rm ext}\left(\pi\right)=\sigma_{\rm sca}\left(\pi\right)+\sigma_{\rm abs}\left(\pi\right)$. Explicitly, the usual integrated quantities are \cite{Gouesbet1988}:

\begin{align}
\sigma_{\rm ext}&=\frac{4\pi}{k^2}\mathfrak{R}\left(\sum_{n,m}N_n\frac{\left(n+|m|\right)!}{\left(n-|m|\right)!}\!\!\left[a_n |g_{n,TM}^m|^2\!+b_n |g_{n,TE}^m|^2\right]\right)\nonumber\\
\sigma_{\rm sca}&=\frac{4\pi}{k^2}\sum_{n,m}N_n\frac{\left(n+|m|\right)!}{\left(n-|m|\right)!} \left[|a_n|^2 |g_{n,TM}^m|^2\!+|b_n|^2 |g_{n,TE}^m|^2\right]\nonumber\\
\sigma_{\rm abs}&=\sigma_{\rm ext}-\sigma_{\rm sca}\label{sigmaAbs}
\end{align}

\section{Exact Beam Shape Coefficients}
The beam shape coefficients (BSCs) describe the incident fields $\mathbf{E}^i$ and $\mathbf{H}^i$. A similar approach as GLMT has been used by A. Neves et al. to model optical trapping \cite{Neves2007}. The expansion coefficients $G_{n,TM/TE}^m$ as given in this reference may be related to the BSCs ($g_{n,TM/TE}^{m}$) used in the framework of the GLMT. 

\begin{align}
\left[g_{n,TM}^{m\ge 0}\atop g_{n,TE}^{m\ge 0}\right] &=\sqrt{\frac{n\left(n+1\right)}{4\pi\left(2n+1\right)}\frac{\left(n-|m|\right)!}{\left(n+|m|\right)!}}\,i^{n+1}\left[-G_{n,TM}^{m}\atop G_{n,TE}^{m}\right]\label{EqnGnmExpl}
\end{align} 

Neves' expansion coefficients describe an aberrated focused beam. It is the resulting field when illuminating a lens of diameter $D=2f\sin\alpha_m$ (the microscope illumination objective, numerical aperture ${\rm NA}_{\rm ill}=n_1\sin\left(\alpha_m\right)$, effective focal length $f$) by a collimated Gaussian beam of width $\omega_a$. The over- / underfilling is then parametrized by the ratio $D/\omega_{\rm a}$ (cf. overfilling-factor $\beta_G=\frac{f}{\omega_{\rm a}}\frac{\rm NA_{\rm ill}}{n_0}$ \cite{Nasse2010}). The BSCs for x-polarized TEM-00 focused laser beam read (for other polarizations, see \cite{Neves2007}):
\begin{align}
\left[G_{n,TM}^{m}\atop G_{n,TE}^{m}\right]& =\pm 4 \pi k_2 f e^{-i k_1 f}i^{n-m+1} e^{-i m \phi_0} \frac{\sqrt{n_0 n_2}}{n_1}\times\label{eqnGnm}\\
&\sqrt{\frac{\left(2n+1\right)\left(n-m\right)!}{4\pi n \left(n+1\right)\left(n+m\right)!}} \,\times I_{n,{TM\atop TE}}^{m}\nonumber\\
I_{n,{TM\atop TE}}^m& =\int_{0}^{\alpha_m}\sqrt{\cos\alpha_1}e^{-f^2\sin^2\alpha_1 /\omega_a^2}\exp\left(i\Psi\right)\times\\
&\left(C_{n,{TM\atop TE}}^m\left[{\cos\phi_0\atop \sin\phi_0}\right]+i m S_{n,{TM\atop TE}}^m\left[{\sin\phi_0\atop -\cos\phi_0}\right]\right)\mathrm{d}\alpha_1\nonumber\\
C_{n,{TM\atop TE}}^m&=\sin\alpha_2\left(m^2 t^{s\atop p}\frac{J_m\left(X_{\alpha_2}\right)}{X_{\alpha_2}}\Pi_n^m + t^{p\atop s}J_m^{'}\left(X_{\alpha_2}\right)\tau_n^{m}\right)\nonumber\\
S_{n,{TM\atop TE}}^m&=\sin\alpha_2\left(t^{s\atop p}J_m^{'}\left(X_{\alpha_2}\right) \Pi_n^m + t^{p\atop s} \frac{J_m\left(X_{\alpha_2}\right)}{X_{\alpha_2}}\tau_n^{m}\right)\nonumber
\end{align}

\noindent where $\rho_0^2=x^2+y^2, \phi_0=\arctan\left(y/x\right)$, $i\Psi= -i k_2 z_0\cos\alpha_2-i\left(k_1\cos\alpha_1-k_2\cos\alpha_2\right)d$, $X_{\alpha_2}=k_2 \rho_0 \sin\alpha_2=k_1 \rho \sin\alpha_1$ and $\cos\alpha_2 = \left(1-\left(n_1/ n_2\right)^2 \sin^2\alpha_1\right)^{1/2}$. Here, $J_m$ are the Bessel functions of the first kind of $m$-th
order and $\Pi_n^m=\Pi_n^m\left(\cos\alpha_2\right)$ and $\tau_n^m=\tau_n^m\left(\cos\alpha_2\right)$ are related to the associated Legendre polynomials (see eqn. (\ref{eqnAngularFunc})). Also the \textsc{Fresnel} transmission coefficients are introduced:
\begin{equation}
t^s=2\left(1+\frac{n_2}{n_1}\frac{\cos\alpha_2}{\cos\alpha_1}\right)^{-1}, \quad t^p=2\left(\frac{n_2}{n_1}+\frac{\cos\alpha_2}{\cos\alpha_1}\right)^{-1}
\end{equation}

\begin{figure}[h]
  \centering
 	\includegraphics[scale=0.7]{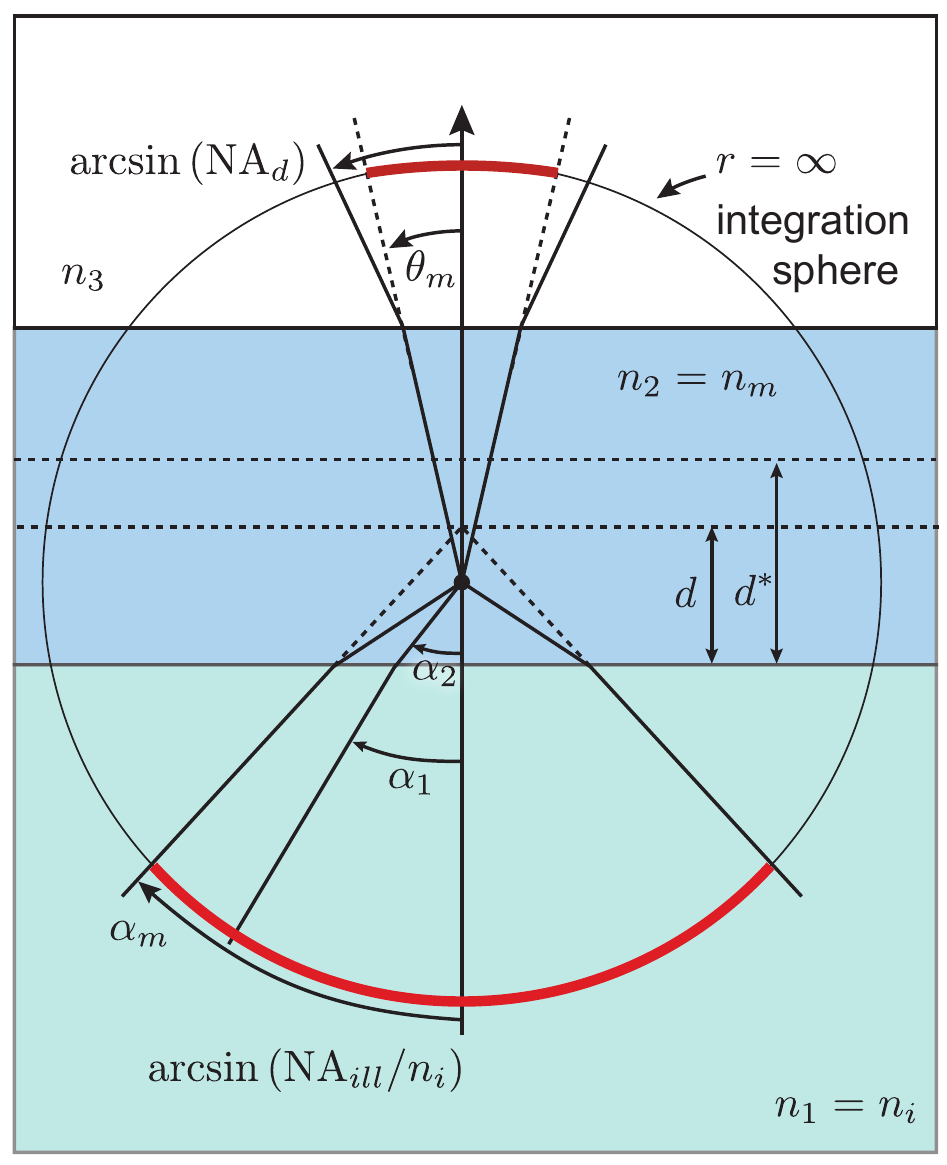}
  \caption{Integration domain of the fluxes, BSC angles}
  \label{fig:Sketch}
\end{figure}

Here, $n_0$, $n_1$ and $n_2$ (see Fig.\ \ref{fig:Sketch}) are the refractive indices of air, immersion-oil/cover-glass and the sample medium. To include the intrinsic objective design aberration which cancels the aberration of a glass-cover-slip under design conditions ($n_0=1$, $n_1=n_g=n_g^{*}$, $d_g=d_g^{*}$ etc.), one may introduce an additional initial aberration function in the integrand of $I_{n,{TM\atop TE}}^m$ as a further phase $\Psi_i^{*}$, i.e. an additional phase factor $\exp (-ik_0\Psi_i^{*})$. To simplify matters, a reasonable form is 
\begin{equation}
\Psi_i^{*}=-d^{*}n_1^{*}\cos\alpha_1^{*},\quad n_1\sin\alpha_1=\sin\alpha_1^{*}n_1^{*}\label{eqnAberrObj}
\end{equation}
motivated by the treatment given by Woehl et al. \cite{Nasse2010}. The parameter $d^{*}$ is a fit-parameter that physically contains all otherwise non-considered aberrations that might be present due to non-design parameters of immersion-oil and coverglass (layer-thicknesses, refractve indices) and the additional interface. This non-considered interface between coverglass and immersion-oil has low optical contrast, since $n_g\approx n_i$ in the experiment. This justifies those subsummations into a single term, and eqn.\ (\ref{eqnGnm}) with $I_{n,{TM\atop TE}}^m$ including the factor $\exp(-ik_0\Psi_i^{*})$ (\ref{eqnAberrObj}) yields upon conversion through eqn.\ (\ref{EqnGnmExpl}) the final form of the exact incident beam expansion BSCs:
\begin{align}
g_{n,{TM\atop TE}}^{m\ge 0} &=\left(-1\right)^n\frac{\left(n-m\right)!}{\left(n+m\right)!} \frac{\sqrt{n_0 n_2}}{n_1} k_2 f e^{-i k_1 f}i^{-m} e^{-i m \phi_0} I_{n,{TM\atop TE}}^{m}\label{eqngnmfinal}
\end{align}
The BSCs need to be computed numerically for positive $m$, since $g_{n,TM,TE}^{-m}=\left(-1\right)^{n+1}g_{n,TM,TE}^{m*}$ for all $m$.

\section{Special Functions\label{secFunc}}
The angular functions appearing within the GLMT framework are:
\begin{align}\Pi_n^m\left(\cos\theta\right)&=\frac{P_n^m\left(\cos\theta\right)}{\sin\theta},&&\tau_n^m\left(\cos\theta\right)=\frac{\mathrm{d}P_n^m\left(\cos\theta\right)}{\mathrm{d}\theta}\label{eqnAngularFunc}\end{align} 

They were derived recursively through the following recurrence-relations, which generalize the expressions given in (\cite{Pena2009}, loc.\ eqn.\ 26(a-c)) and add some trivial identities obtained from known identities for the associated Legendre polynomials $P_n^m$:
\begin{align}
\Pi_{0}^{0}&=1/|\sin\theta|\\
\Pi_{n}^{n}&=\left(-1\right)^n \frac{\left(2n\right)!}{2^n n!}\sin^{n-1}\theta\\
\Pi_{n}^{n-1}&=\cos\theta\left(2n-1\right)\Pi_{n-1}^{n-1}\\
\Pi_{n}^{m}&=\frac{2n-1}{n-m}\cos\theta \,\Pi_{n-1}^{m}-\frac{n+m-1}{n-m}\Pi_{n-2}^{m}
\end{align}

The fourth relation is only to be used for $m\le n-2$. For the derivatives of the associated Legendre polynomials, $\tau_n^m$, one may use the following recurrence relations afterwards:
\begin{align}
\tau_0^0&=0\\
\tau_n^{n}&=-n\left(2n-1\right)\sin\theta\cos\theta \,\Pi_{n-1}^{n-1}\\
\tau_n^{m}&=n\cos\theta \,\Pi_n^{m}-\left(n+m\right)\,\Pi_{n-1}^{m}\label{eqnAngleLast}
\end{align}

Also the following relation for the derivatives of the Bessel-functions were used:
\begin{align}
J'_m\left(x\right)=\frac{\mathrm{d}J_m\left(x\right)}{\mathrm{d}x}&=\left[J_{m-1}\left(x\right) - J_{m+1}\left(x\right)\right]/2
\end{align}

\section{Incident and Absorbed Powers}
Per solid angle $\mathrm{d}\Omega=\sin\theta\mathrm{d}\phi\mathrm{d}\theta$ the power $\mathrm{d}P=I\left(\rho\right)\rho\mathrm{d}\rho\mathrm{d}\phi$ is contained. The \textsc{Abbe} sine condition relates the incident beam cylinder sector distance $\rho$ to the angle $\theta$ via the effective focal length $f$ of the objective: $\rho=f\sin\theta$. Therefore, $\mathrm{d}r/\mathrm{d}\theta=f\cos\theta$ and the intensity is obtained from the Gaussian incident intensity profile $I\left(\rho\right)=I_0\exp\left(-2\rho^2/\omega_a^2\right)$ with beam waist $\omega_a$ and the definition of an overfilling factor $\gamma=f/\omega_a$:

\begin{align}
P_{\rm inc}&=\int_0^{2\pi}\int_{0}^{r_{\rm obj,e}}I\left(\rho\right)2\pi \rho \mathrm{d}\rho\mathrm{d}\phi\nonumber\\
 &=2\pi\int_0^{\alpha_m}f^2 \cos\theta\sin\theta I_0\exp\left(-2\gamma^2 \sin^2\theta\right)\mathrm{d}\theta\nonumber\\
 &=I_0\frac{\pi f^2}{2\gamma^2}\left[1-\exp\left(-2\gamma^2\sin^2\theta_m\right)\right]\label{eqnsigmainc}
\end{align}

here and later the subscript $_e$ denotes the reference to the physical back entrance aperture of the objective, and $r_{\rm obj, e}$ is the radius of the back aperture. The expression goes to the plane-wave(pw)-limit of $P^{\rm pw}_{\rm inc}=I_0\pi f^2\sin^2\alpha_m$ for $\gamma \ll 1$ with constant illumination intensity $I_0$ filling the aperture. This limit was considered in \cite{Lerme2008}. The general expression (\ref{eqnsigmainc}) may be used to obtain an estimate of the detected power without a particle, i.e.\ for normalization (see next section). For heating estimates, the incident power on the particle is needed, and hence $\alpha_m=\arcsin\left({\rm NA}_{\rm ill}/n_1\right)$ is taken, while for the detection background intensity $\alpha_m=\theta_m=\arcsin\left({\rm NA}_d/n_2\right)$ has to be used (assuming $\theta_m < \alpha_m$). Analogously to the cross-sections, we may define the following two quantities:
\begin{align}
\sigma_{\rm inc}^{\rm ill}&=P^{\rm ill}_{\rm inc}/I_0, \qquad &\alpha_m&=\arcsin\left({\rm NA}_{\rm ill}/n_1\right)\\
\sigma_{\rm inc}^{d}&=P^{d}_{\rm inc}/I_0, \qquad &\alpha_m&=\arcsin\left({\rm NA}_d/n_2\right)\label{eqnSigmadExact}
\end{align}

Also, $\sigma_{\rm abs}=P_{\rm abs}/I_0$ and thus $\sigma_{\rm abs}/\sigma^{\rm ill}_{\rm inc}=P_{\rm abs}/P^{\rm ill}_{\rm inc}$ which may be used to calculate absorbed powers via 

\begin{equation}
P_{\rm abs}=P^{\rm ill}_{\rm inc} \frac{\sigma_{\rm abs}}{\sigma^{\rm ill}_{\rm inc}}
\end{equation}

In case of a power-meter detection area radius $r_{\rm PM}$ deviating from the objective back-aperture ($r_{\rm obj,e}$), a further factor to be considered is:

\begin{align}
c_{\rm PM}=&\frac{P^{\rm ill}_{\rm inc}}{P^{\rm PM}_{\rm inc}}=\int_{0}^{r_{\rm obj,e}}\!\!\!\!\!\!I\left(\rho\right)2\pi \rho \mathrm{d}\rho \,\Biggr{/}\, \int_{0}^{r_{\rm PM}}\!\!\!\!\!\!I\left(\rho\right)2\pi \rho \mathrm{d}\rho\nonumber\\
&=\frac{1-\exp\left(-2 r^2_{\rm obj,e}/\omega^2_{\rm a,e}\right)}{1-\exp\left(-2 r^2_{\rm PM}/\omega_{\rm a,e}^2\right)}
\end{align}

Given an aperture radius $r_{\rm obj,e}$ and beam-waist $\omega_{a,e}$ of the incident beam we may relate these parameters to the effective lens parameters via $r_{\rm obj}=f \sin\theta_m=f {\rm NA_{\rm ill}}/n_1$ and $r_{\rm obj,e}/\omega_{\rm a,e}=r_{\rm obj}/\omega_{\rm a}$, such that $\gamma =\frac{r_{\rm obj,e}}{\omega_{\rm a,e}} \frac{n_1}{{\rm NA}_{\rm ill}}$. A further factor of $c_{T,h}=0.86$ or $c_{T,d}=0.83$ for $532$ nm (subscript $h$) or $635$ nm (supscript $d$) wavelength accounts for the transmission of the used objective (Olympus, UPlanSApo 100XO).
Heating may thus be computed via
\begin{align}
\Delta T_0&=\frac{\sigma^{\lambda_h}_{\rm abs}}{\sigma^{\rm ill}_{\rm inc}} c_{T,\lambda_h} c_{\rm PM} \frac{P_{\rm PM}}{4\pi \kappa R}.
\end{align}

To connect the exact with an approximate Gaussian calculation we note the following. The absorbed power $P_{\rm abs}=I_0 \sigma^{G}_{\rm abs}$ may be obtained with the help of the gaussian GLMT absorption cross-section $\sigma^G_{\rm abs}$. The total power $P^{\rm total}_{\rm inc}$, which is to be used in the Gaussian approximation in order to compute the peak intensity via $I_0=2P^{\rm total}_{\rm inc}/\left(\pi \omega_0^2\right)$, reads
\begin{align}
P^{\rm total}_{\rm inc}=c_{T,\lambda_h} c_G c_{\rm aberr} c_{\rm PM}P^{\rm PM}_{\rm inc},
\end{align}

wherein the factors $c_G$, $c_{\rm aberr}$ have the following meaning:
\begin{align}
c_{G}=&\frac{P^{\rm total}_{\rm inc}}{P^{\rm obj}_{\rm inc}}=\int_{0}^{\infty}\!\!\!\!\!\!I\left(\rho\right)2\pi \rho \mathrm{d}\rho \,\Biggr{/}\, \int_{0}^{r_{\rm obj,e}}\!\!\!\!\!\!I\left(\rho\right)2\pi \rho \mathrm{d}\rho\label{eqncG}\\
=&\left[1-\exp\left(\frac{-2 r_{\rm obj,e}^2}{\omega_{a,e}^2}\right)\right]^{-1}=\left[1-\exp\left(\frac{-2 \gamma^2 {\rm NA}^2}{n_1^2}\right)\right]^{-1}\nonumber.
\end{align}

For $\gamma=1$ and $n_1=1.46$ and ${\rm NA}_{\rm ill}=1.4$ the coefficient takes the value $c_G=1.225$. Aberrations may be included by introducing an intensity scaling factor. This factor may be obtained by considering in the exact GLMT $z_p$-scan of $\sigma^E_{\rm abs}$ for the aberrated case and for the un-aberrated case and considering the peak-ratio. Alternatively, it may be equally well obtained by a PSF analysis according to \cite{Nasse2010}.
\begin{equation}
c_{\rm aberr}=\frac{{\rm max}\left(P_{\rm abs}^{\rm aberr}\right)}{{\rm max}\left(P_{\rm abs}^{\rm unaberr}\right)}= \frac{{\rm max}\left(\sigma_{\rm abs}^{\rm E, aberr}\right)}{{\rm max}\left(\sigma_{\rm abs}^{\rm E, unaberr}\right)}\label{eqncAberr}
\end{equation}

This means, that the following absorbed powers can be compared and used to calculate the temperature rise induced via $\Delta T_0=P_{\rm abs}/\left(4\pi \kappa R\right)$:

\begin{align}
\textnormal{Gaussian:}&\quad P_{\rm abs}^G=c_{T,\lambda_h} c_G c_{\rm aberr} \frac{2 c_{\rm PM} P_{\rm PM}}{\pi \omega_0^2} \sigma_{\rm abs}^{G}\label{eqnPabsG}\\
\textnormal{Exact:}&\quad P_{\rm abs}^E=c_{T,\lambda_h} c_{\rm PM} P_{\rm PM} \frac{\sigma_{\rm abs}^E}{\sigma^{\rm ill}_{\rm inc}}\label{PabsExact}
\end{align}

\section{The Photothermal Signal}
A normalization is needed in order to define the relative photothermal signal (eqn.\ \ref{eqnRelPTSignal}). A quantity having the units of an area and being proportional to the background power has been introduced for that purpose: $\sigma_{\rm inc}^{d}$. For the exact beam, $\sigma_{\rm inc}^{d}$ from eqn.\ (\ref{eqnSigmadExact}) and (\ref{eqnsigmainc}) should be used. For a Gaussian beam (on-axis) one may calculate the integrated flux of the collected beam directly, i.e. $\int\mathbf{S}^{i} \cdot \mathrm{d}\mathbf{a}$ with $2S_{\perp}^i=\mathfrak{R}\left[E_\theta^i H_\phi^{i*}-E_\phi^i H_\theta^{i*}\right]/2$. The result may be written as a Cauchy-sum $\sigma_{\rm inc}^d= \frac{\pi}{2k^2}\sum_{n=1}^{\infty} \sigma_{{\rm inc},n}^d$ with summands $\sigma_{{\rm inc},n}^d$ given by:
\begin{align}
\sigma_{{\rm inc},n}^d=&\sum_{m=1}^{n}N_m g_m N_{n-m+1}g_{n-m+1}^{*}\int_{0}^{\theta_m}\\
&\left[\Sigma_m\Sigma_{n-m+1}-\left(-1\right)^{n}\Delta_m\Delta_{n-m+1}\right]\sin\theta\mathrm{d}\theta\nonumber
\end{align}

with $\Delta_n\equiv \Pi_n- \tau_n$ and $\Sigma_n\equiv \Pi_n+ \tau_n$. To ensure numerical stability for small angles, a direct recursive determination of $\Delta_n\equiv \Pi_n- \tau_n$ was used \cite{Meeten1984}:

\begin{align}
\Delta_0=&\,0, \,\,\,\,\, \Delta_1=1-\cos\theta, \,\,\,\,\,\Delta_2=3+3\cos\theta-6\cos^2\theta\\
\Delta_n=&\frac{2n-1}{\left(n-1\right)^3}\left[1+n\left(n-1\right)\cos\theta\right]\Delta_{n-1} - \frac{n^3}{\left(n-1\right)^3}\Delta_{n-2}
\end{align}

The relative photothermal signal may be calculated quantitatively and compared to experimental values by considering relative signals, i.e.\ the ratio of the time-varying power, which is the photothermal signal $S$, and the large constant power $P_{\rm PD}$ detected i.e. on a photodiode (voltage $V_{\rm PD}$):

\begin{align}
\Phi& =\frac{\Delta P_{\textnormal{PD}}}{P_{\textnormal{PD}}}=\frac{P_{\textnormal{PD, hot}}-P_{\textnormal{PD, cold}}}{P_{\textnormal{PD, cold}}}\approx \frac{\Delta \sigma_{\rm ext} +\Delta \sigma_{\rm sca}}{\sigma_{\rm inc}^d}\\
&= \frac{\left(\sigma_{\rm sca}+\sigma_{\rm ext}\right)|_{\Delta T_0\left(P_h,P_d\right)} - \left(\sigma_{\rm sca}+\sigma_{\rm ext}\right)|_{\Delta T_0\left(P_h=0,P_d\right)}}{\left.\sigma_{\rm inc}^d\right|_{z_p=0}}\label{eqnRelPTSignal}
\end{align}

In a modulated heating laser beam usually the effective heating power $P_{PM,\textnormal{eff}}=P_{PM} / 2$ is detected by a power-meter ($P_{\rm PM}$ being the peak-to-peak power used for calculations and static experiments), and furthermore a lock-in amplifier that is common to these experiments outputs a photothermal (rms, root-mean-squared) voltage $V_{\rm PT}^{\rm rms}$ that is related to the peak-to-peak voltage via $V_{{\rm PT}}^{\rm rms}=V_{{\rm PT}}^{\rm pp}/\left(2\sqrt{2}\right)$. Therefore,

\begin{equation}
\Phi=\left.\frac{\Delta P_{\rm PD}}{P_{\rm PD, cold}}\right|_{P_{\rm PM}}  =2\sqrt{2} \frac{V_{\rm PT}^{\rm rms}}{V_{\rm PD}}.
\end{equation}

\section{Gaussian Beam Shape Coefficients\label{GaussGLMT}}
While the implementation of an off-axis beam according to \cite{Ren1992} may be used to generate images akin to Fig.\ 1 of the article, the aberration-caused details are naturally absent in these calculations. Considerably easier than the off-axis scattering is the computation of the signal for a particle illuminated on-axis by a focused Gaussian beam. In that case, the double and triple sums simplify and one may utilize the modified local approximation (MLA, \cite{Gousebet1995modlocApprox}):
\begin{equation}g_n\left(s,\gamma\right)=\overline{Q}\exp\left(-\overline{Q}s^2\left(n-1\right)\left(n+2\right)\right)\exp\left(i \gamma s^{-1}/2\right)\label{eqngn},\end{equation} 

where $\overline{Q}=\left(1+i s \gamma\right)^{-1}$ with the beam-confinement factor $s$ defined through $s=\omega_0/\left(2z_R\right)=1/k \omega_0$, i.e. the ratio of lateral to axial extent of the beam focus, and a defocussing parameter $\gamma=2 z_p/\omega_0$. $z_p$ describes the axial position of the beam-focus relative to the center of the scatterer, $z_p$ being the displacement of the particle relative to the beam-waist: $z_p <0$ corresponds to the situation where the focus is between the particle and the collecting objective. The \textsc{Rayleigh}-range of the Gaussian beam is $z_R=n_{m}\pi \omega_0^2/\lambda$ and the beam waist is $\omega_0$. The wave-vector is given by $k=2\pi n_{m}/\lambda$ with $n_{m}$ being the refractive index of the particle-embedding medium far away from it. The results expressed as cross-section in any forward/backward angular domain are:
\begin{align}
\sigma_{\rm sca}\left(\theta_m\right)=&\, \frac{\pi}{k^2}\int_{0}^{\theta_m}\left(|S_{1}\left(\theta\right)|^2+|S_{2}\left(\theta\right)|^2\right)\sin\theta \mathrm{d}\theta\label{eqnScatter}\\
\sigma_{\rm ext}\left(\theta_m\right)=&\frac{-\pi}{k^2}\int_{0}^{\theta_m}\left[\mathfrak{R}\left(M\right)\,\mathfrak{R}\left(S_1+S_2\right)\right.+\nonumber\\ &\qquad\qquad\qquad \left. \mathfrak{I}\left(M\right)\,\mathfrak{I} \left(S_1+S_2\right)\right]\sin\theta \mathrm{d}\theta\label{eqnISI}\\
M\left(\theta\right)=&\sum_{n=1}^{\infty}\,\left[\Pi_n\left(\cos\theta\right)+\tau_n\left(\cos\theta\right)\right]\,g_n N_n
\end{align}

The usual full-$\pi$-integrated quantities are
\begin{align}
\sigma_{\rm sca}&=\frac{2\pi}{k^2}\sum_{n=1}^{\infty}\left(2n+1\right) |g_n|^2 \left(|a_n|^2+|b_n|^2\right)\label{SigmaExt}\\
\sigma_{\rm ext}&=\frac{2\pi}{k^2}\sum_{n=1}^{\infty}\left(2n+1\right) |g_n|^2 \,\mathfrak{R}\,\left(a_n+b_n\right)\\
\sigma_{\rm abs}&=\sigma_{\rm ext}-\sigma_{\rm sca}.\label{OnAxissigmaAbs}
\end{align}

\bibliographystyle{plain}
%\bibliography{MyLibrary9}
%\bibliography{MyLibraryAppendix}